\documentclass{article}
\usepackage{sao2}
\usepackage{psfig}
\issue{2001, 51, 11-20  }

\title{Measurement of radial velocities and velocity dispersion of stars in
circumnuclear regions of galaxies using the 2D spectrosopy technique}
\author{A.V. Moiseev}
\institute{\saoname}
\date{February 12, 2001}{February 19, 2001}
\begin{document}
\maketitle
\begin{abstract}
A procedure is described of construction of velocity fields and velocity
dispersion of the stellar component in the central regions of galaxies on
the basis of data  obtained with the integral field spectrograph MPFS.
Cross-correlation  and Fourier quotients techniques adapted for 2D spectra
reduction are used. A detailed discussion is given of the problems of
providing for the instrumental profile in measuring the velocity dispersion.
Using the existing version of the spectrograph, one can measure radial
velocities and the dispersion of velocities to an accuracy of 5--10\,km/s and 10--15\,km/s,
respectively. As an example the ``double-barred'' galaxy NGC\,2950 is
considered.
\keywords{galaxies: fundamental parameters --- galaxies: individual: NGC\,2950
--- galaxies: spiral}
\end{abstract}

\section{Introduction}
The present paper examines the problem of determination of two basic observational
parameters that characterize the collective motion of stars in galaxies ---
radial velocity $v$ and velocity dispersion $\sigma$. The task has
a half-a-century history, beginning with the pioneering observations by
Minkowski (1954) to the latest papers concerning the relation between the
mass of a central black hole and the velocity dispersion (Gebhardt et al.,
2000). The procedure of measuring $v$ and $\sigma$ with the application
of the technique of Fourier transform goes back to the classical works
of Tonry and Davis (1979, hereafter TD) and Sargent et al. (1977,
hereafter SSBS).

The methods of two-dimensional (2D) spectroscopy that have been advanced
intensively over the last decade have made it possible to obtain
two-dimensional mapping of velocities and velocity dispersions. In the
case of disk galaxies this enables one to study non-circular motions in the
region of the triaxial potential (bar) or near the active nucleus
(see the survey of Arribas and Mediavilla, 2000). The multipupil spectrograph
(MPFS) has been in operation at SAO RAS since 1990. It has been used to
conduct similar observations (see, for instance, Afanasiev et al., 1996;
Sil'chenko, 1998). In these papers, however, field velocities of stars and
gas alone have been discussed. At the same time, the distribution of
velocity dispersion gives  information about the non-axisymmetrical shape
of the gravitational potential. This is one of the key parameters in
producing numerical dynamical models of galaxies (Khoperskov and Chulanova,
2001). The main methodological problem in measuring velocity dispersions
is taking correct account of the variations of the instrumental profile of the
spectrograph. A new version of the MPFS (in operation at BTA since 1998) constructs
reliably the distributions of velocity dispersions in circumnuclear
regions of galaxies, which is shown in the present paper.

The paper is organized as follows. The theory of methods of estimating
$v$ and $\sigma$ is considered in Section 2. The problems of provision
for the instrumental profile are discussed in Section 3, Section 4 gives a
brief description of algorithms of analysis of MPFS spectra and errors
that arise, the results of the galaxy NGC\,2950 are presented in
Section 5, conclusive remarks are collected in Section 6.

\begin{figure}
\psfig{figure=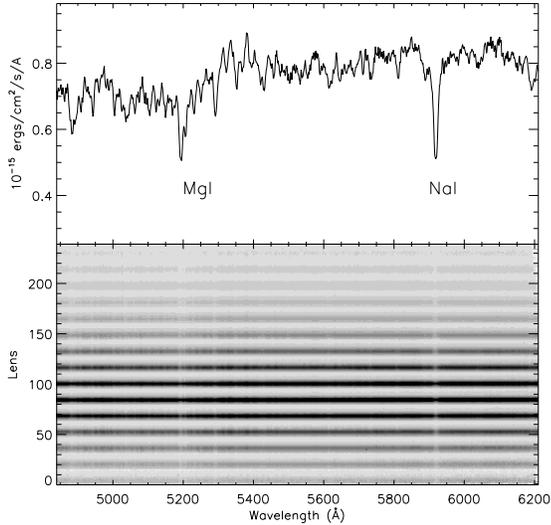,width=8cm}
\caption{Example of MPFS spectra: the 2D spectrum of the galaxy NGC\,2950
(bottom); the spectrum corresponding to the galaxy centre (top).}
\label{mpfs2950}
\end{figure}
\section{Methods of velocity and velocity dispersion measurement}
\subsection{General remarks}
After a preliminary reduction, the MPFS observational data can be presented
as two-dimensional spectra. The spectra from different spatial elements
are brought to a common wavelength scale and arranged in the frame
sequentially one above another. An example of such spectra is displayed in
Fig.\ref{mpfs2950}. The spatial elements are constructed as small lenses and
grouped into an array of $16\times15$ pixels (micropupils). This corresponds
to a field of view of $16''\times15''$ on the sky.

\begin{figure}
\psfig{figure=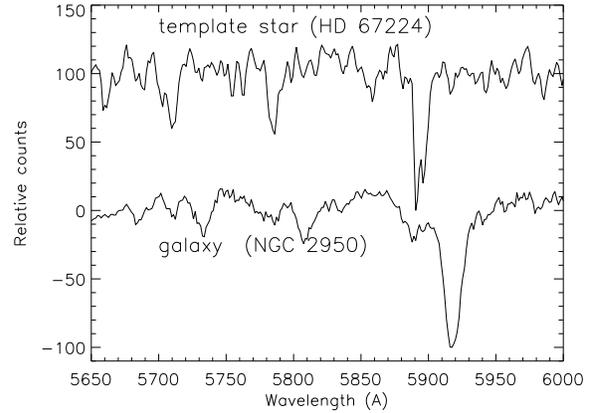,width=8 cm }
\caption{Spectra of the template star and of the S0 galaxy nucleus in the
region of the doublet Na\,I obtained with the MPFS. The wavelength
displacement of the galaxy spectrum is due to the Doppler shift, the line
broadening is caused by the  velocity dispersion along the line of
sight.}
\label{gal_str1}
\end{figure}

Let us illustrate the idea of the methods of measuring $v$ and $\sigma$ by
an individual spectrum of one spatial element. Let $G(\lambda)$ and $T(\lambda)$
are the observed spectra of the galaxy and a template star with the
low-frequency component (continuum) subtracted. Typical examples of such
spectra are shown in Fig.\,\ref{gal_str1}.
The spectra are discretely sampled in wavelength $\lambda$.

The observed spectra need to be converted to a logarithmic scale in which
the number of the spectral bin $n$ is related to the wavelength $\lambda$
by the expression
\begin{equation}
 n=a\log(\lambda)+b,
\label{ton}
\end{equation}
where
\begin{equation}
 \begin{array}{l}
a=\log^{-1}(1+\Delta v/c), \\
 b=-a\,\log(\lambda_0), \\
  \end{array}
\end{equation}
here $\Delta v$ is the step in velocity expressed in km/s, $c$ is the
speed of light. The Doppler shift leads to uniform shift of the spectrum
along the $n$ axis, irrespective of $\lambda$.

The galaxy spectrum may be approximately represented as a redshifted
template spectrum convolved with a certain broadening function caused by
inner motions of stars in the galaxy along the line of sight:
\begin{equation}
  G(n)\approx \alpha T(n)\odot B(n-v),
\label{main}
\end{equation}
the broadening function $B(n)$ in a first approximation can be represented
by a Gaussian:
\begin{equation}
  B(n)=\frac{1}{\sqrt{2\pi}\sigma}\exp\left(-\frac{n^2}{2\sigma^2}\right),
\label{broad}
\end{equation}
where  $\alpha$ is the numerical coefficient, $v$ and $\sigma$ are the
velocity and velocity dispersion along the line of sight.

The methods described below are based on calculations of a discrete
Fourier transform defined for the function $F(n)$ as
\begin{equation}
\tilde F(k)=\sum_{n=0}^{N-1}F(n)\exp\left(-\frac{2\pi i n k}{N}\right),
\label{fur}
\end{equation}
where $N$ is the number of points in the spectrum.

\subsection{Cross-correlation method}
The TD method is based on computation of a normalized cross-correlation
function determined as
\begin{equation}
\begin{array}{l}
C(n)=G(n)\otimes T(n)=\\ \qquad
{\displaystyle \frac{1}{N\sigma_g\sigma_t}\sum_m G(m)T(m-n)},\\
 \end{array}
\label{cor}
\end{equation}
where
$$
\sigma_g^2=\frac{1}{N}\sum_{n}G(n)^2,\,
\sigma_t^2=\frac{1}{N}\sum_{n}T(n)^2.
$$

After Fourier transformation, relationship (\ref{cor}) is written as
\begin{equation}
\tilde C(k)=\frac{1}{N\sigma_g\sigma_t}\tilde G(k)\tilde T^*(k).
\end{equation}
where stars marks the complex conjugation.

Applying the procedure of cross-correlation to expression  (\ref{main}),
one can readily obtain
\begin{equation}
C(n)=(T(n)\otimes T(n))\odot B(n-v),
\label{convol}
\end{equation}
i.e. the cross-correlation of the galaxy and star spectra is a convolution
of the auto-correlation of the template star spectrum $T(n)$ (which
provides information about the instrumental profile of the spectrograph)
with the broadening function $B(n)$. The velocity $v$ is thus defined
from the location of the cross-correlation function $C(n)$ maximum.

\begin{figure*}
\psfig{figure=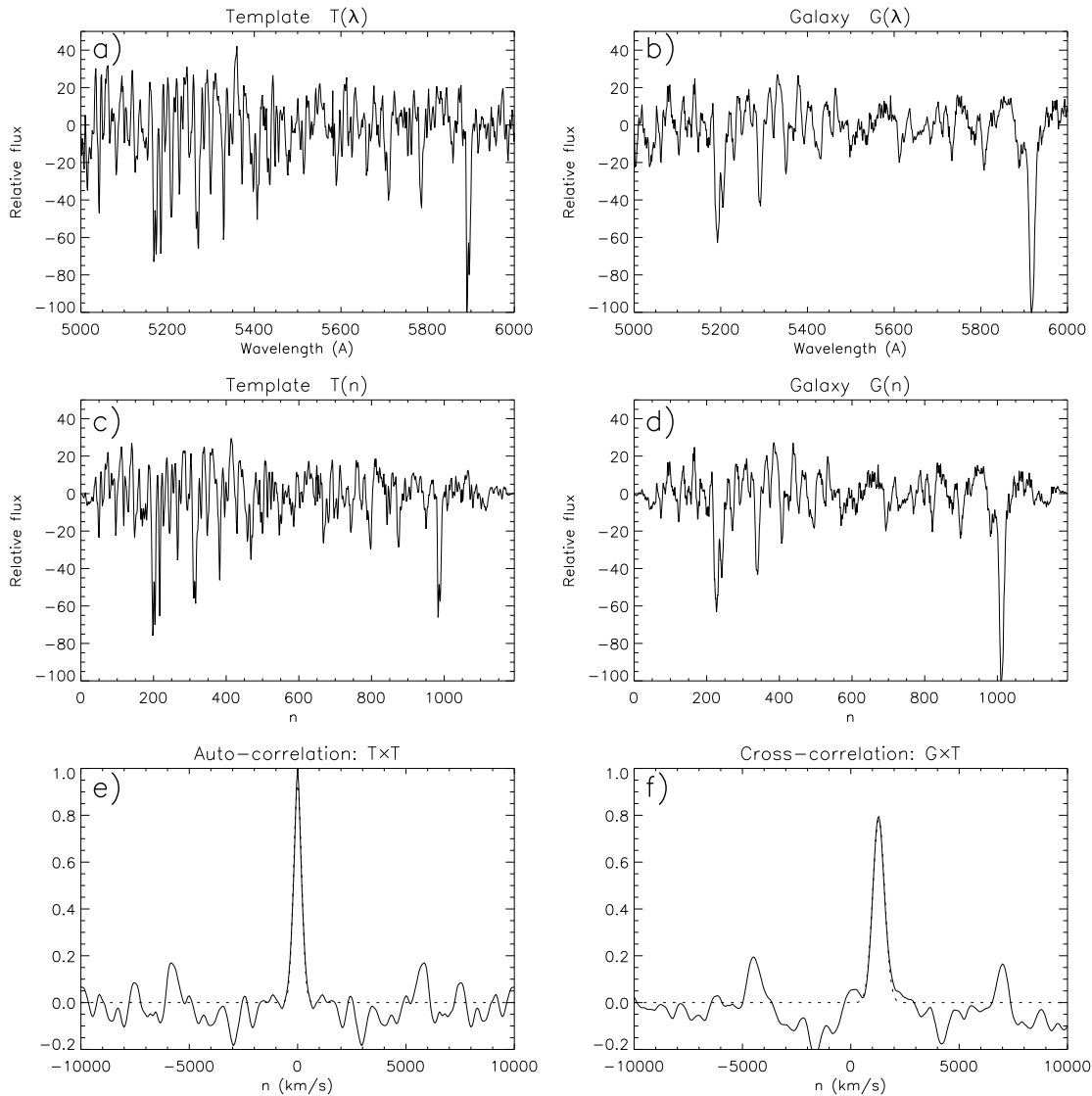,width=15 cm }
\caption{Cross-correlation method. a, b are the spectra of a K\.III star
and the galaxy NGC\,2950, the continuum subtracted; c, d are the spectra
on a logarithmic scale with a step of $50\,km/s$, the boundaries of the
spectra are multiplied by a sinusoid (see 4.1);
e, f are the auto-correlation and the cross-correlation functions,
the dotted line shows the approximation by a Gaussian.}
\label{cor2}
\end{figure*}

If the central peak of the cross-correlation function may be represented
by a Gaussian with a dispersion $\mu$:
\begin{equation}
  C(n)=c_0\exp\left(-\frac{(n-v)^2}{2\mu^2}\right), \\
\label{corv}
\end{equation}
and the peak of the auto-correlation function of the template spectra by
a Gaussian with a dispersion $\sqrt{2}\,\tau$:
\begin{equation}
  T(n)\otimes T(n)=t_0\exp\left(-\frac{n^2}{2(2\tau^2)}\right),
\label{auto}
\end{equation}
then the square of the velocity dispersion is computed  as
\begin{equation}
\sigma^2=\mu^2-2\tau^2.
\label{td}
\end{equation}
In Fig.\,\ref{cor2} are presented examples of the star and galaxy spectra
prepared for cross-correlation, as well as the auto-correlation and
cross-correlation peaks. In practice, due to the difference between the
shape of the peak and the Gaussian one, to the contribution of the
low-frequency component etc., the relation between $\sigma$ and $\mu$
may differ from (\ref{td}), which introduces systematic errors into the
estimate of $\sigma$.

\begin{figure}
\psfig{figure=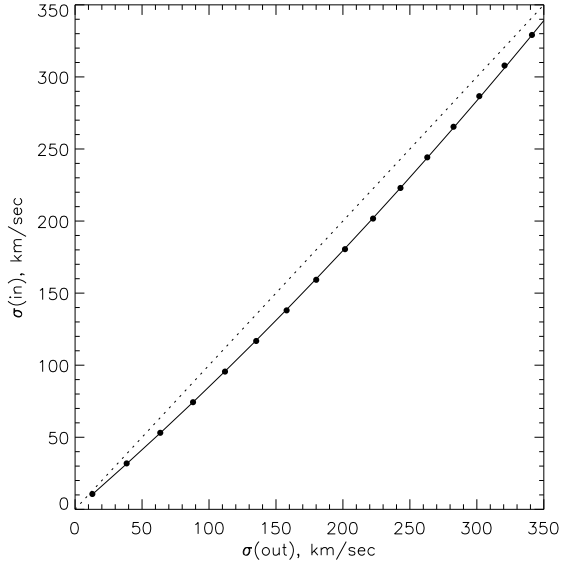,width=8cm }
\caption{Relationship (12): $\sigma_{in}$ is the width of the smoothing
Gaussian, $\sigma_{out}$ is the dispersion estimated from the cross-correlation
peak width (dots). The solid line is the approximation by a second-order
polynomial, the dotted line corresponds to the relation
$\sigma_{out}$\,=\,$\sigma_{in}$.}
\label{calib}
\end{figure}

A number of authors have proposed different modifications of the TD
method for estimating the velocity dispersion. So, in the paper by
Bottema (1988), straightforward use of relation (\ref{convol}) is made ---
the template auto-correlation function is convolved with Gaussians of
different width until the cross-correlation peak shape is best reproduced.
We dwelled on the application of the empirical relation between the
cross-correlation peak width and the velocity dispersion (see Nelson and
Whittle, 1995). For this the template spectrum $T$ is convolved with
Gaussians of different width $\sigma_{in}$, a cross-correlation function
of broadened spectrum with $T$ is constructed, and using formula (\ref{td})
the velocity dispersion
$\sigma_{out}$ is formally determined. The relationship between
$\sigma_{in}$ and $\sigma_{out}$, presented in Fig.\,\ref{calib},
is approximated by a second-order polynomial:
\begin{equation}
 \sigma_{in}=\sum_{i=0}^2\alpha_i\sigma_{out}^i,
\label{calibe}
\end{equation}
and is next used to estimate the dispersion from the cross-correlation
peak width:
\begin{equation}
 \sigma=\sum_{i=0}^2\alpha_i(\mu^2-2\tau^2)^{i/2}.
\label{calibe2}
\end{equation}

\subsection{The Fourier quotient method}
This method was put forward in the SSBS paper. By applying Fourier transform
(\ref{fur}) to (\ref{main}) with provision for  (\ref{broad}), obtain
\begin{equation}
\frac{\tilde G(k)}{\tilde T(k)}=\gamma\exp\left[
-\frac{1}{2}\left(\frac{2\pi k \sigma}{N}\right)^2+\left(\frac{2\pi i k
v}{N}\right)\right],
\label{ssbs}
\end{equation}
the real part in the exponent is defined by the velocity dispersion, the
imaginary part --- by the galaxy velocity. From relation (\ref{ssbs}) both
the parameters  ($\sigma$ and $v$) are formally determined, however,
in practice, the ratio of the Fourier spectra of the object and template is more
noisy than the cross-correlation function peak in the TD method. For this
reason, it is simpler and more reliable to find $v$ from the shift of
the cross-correlation peak; in this case the dispersion is estimated from:
\begin{equation}
\mbox{Re}
\frac{\tilde G(k)\exp(-\frac{2\pi i k v}{N})}
{\tilde T(k)}=\gamma\exp\left[
-\frac{1}{2}\left(\frac{2\pi k \sigma}{N}\right)^2 \right].
\label{ssbs2}
\end{equation}

An example of approximation of the right side of
(\ref{ssbs2}) is shown in Fig.\,\ref{ssbsfig}.  It is necessary to choose
the boundaries in frequency $k$, within which the fitting $k_1<k<k_2$
is made, in order to avoid the influence of high- and low-frequency
noises. Our experience  shows that
$k_1=5-10$ and $k_2=50-120$ are optimum values
for the number of points in the spectrum  $N=500-1000$,
being in good agreement with SSBS.

The advantage of the Fourier quotient method as compared to the
cross-correlation one is that to estimate the dispersion in the
SSBS-technique, no additional constraints on the shape of the
cross-correlation and auto-correlation peaks are required, and only the
assumption that broadening function (\ref{broad}) is of Gaussian shape
is applied.

\section{Instrumental profile problem}
Expression (\ref{auto}) shows the fact that if the spectrograph
instrumental profile is described by a Gaussian with a dispersion $\tau$,
the width of the auto-correlation function of the spectrum produced is
then  $\sqrt{2}\,\tau$ (provided that the lines in the template spectrum
can be considered to be infinitely narrow with respect to  $\tau$). The
width (and even the shape) of the instrumental profile of the spectrograph
changes over the field (along the spectrograph slit and over the
multipupil array), which is caused first of all by the spectrograph
optics, precision of its adjustment and focusing. Besides, due to
temperature fluctuations and flexures of the spectrograph, the quantity
$\tau$ at each point is a function of time of observations and zenith
distance.

Analysis of spectra taken with the new version of the MPFS shows that the
spectrograph instrumental profile is time-stable, i.e. the variations of
$\tau$ during the observing night can be neglected. In this case, the
profile width is defined only by the coordinates of the spatial element a
given spectrum corresponds to: $\tau=\tau(x,y)$.

\begin{figure*}
\psfig{figure=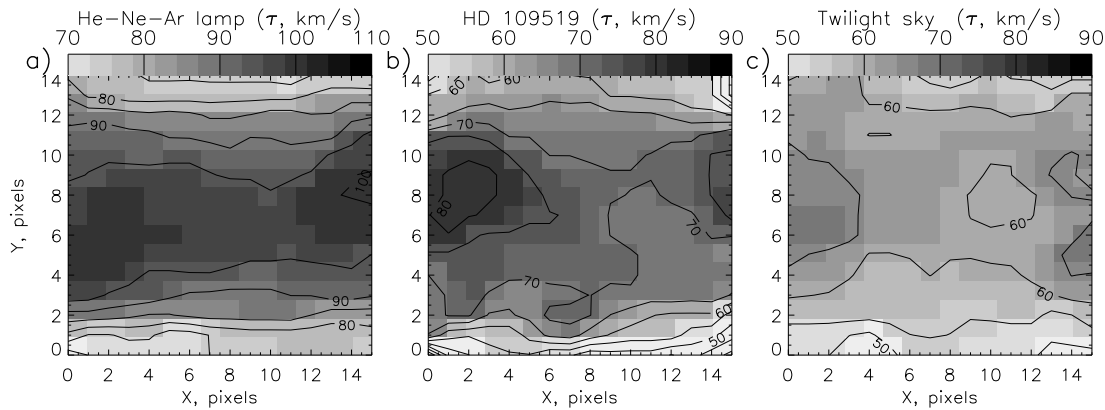,width=15 cm }
     \caption{Instrumental profile of the MPFS (dispersion of the Gaussian
$\tau$ in $km/s$) evaluated in the spectral range 5100--6100\,\AA, the
spectrograph dispersion is
$1.35$\,\AA/px: a --- from the comparison spectrum, b --- from the spectrum
of the KIII star, c --- from the twilight spectrum.}
\label{fwhm}
\end{figure*}

Fig.\,\ref{fwhm} presents estimates of the MPFS profile width for different
spectra (a He-Ne-Ar lamp comparison spectrum, field-of-view defocused
template star spectrum, twilight sky spectrum). The measurements made from
the comparison spectrum are systematically larger by 20--30\,km/s (about
30\% of the mean). This is connected with the fact that the entrance pupil
image for the calibration beam is displaced with respect to the pupil
position when observing the object. It can be seen from Fig.\,\ref{fwhm}
that the profile width variations are $\pm(10\div20)\%$ of the mean
value\footnote{Note that all the measurements of the instrumental profile
are made in terms of dispersion of the Gaussian. The determination of the
profile width as full width on half maximum $FWHM\approx2.355\tau$ is more
usual.} ($\tau\approx70$\,km/s).

We performed experiments using different methods of taking account of the
systematic variations of the instrumental profile with the aim of
reliable mapping of the velocity dispersion distribution $\sigma$ and came
to a conclusion that the following procedure is the most correct and simple
at the same time. A bright ($5^m-6^m$) template star is defocused over
the whole array of micropupils ($16''\times15''$). During moderate
exposures (90--200\,s) at the MPFS it is possible to take template spectra
with high signal-to-noise ratio over the
whole field. When analysing the galaxy spectrum by the TD and SSBS at each
point of $(x, y)$, an individual template spectrum $T(x,\,y)$ is taken.
Such a procedure takes actually full account of the effect of the
instrumental profile, provided, of course, that the spectrograph is
carefully focused before the observation.

\section{Sequence of work with MPFS data}
\subsection{Preparation of spectra}
Using the language IDL, we have written a package of programmes CROSSLIB
which implements the methods presented in Sections 2 and 3. The initial
data are  2D spectra of a galaxy and a star (Fig.\,\ref{mpfs2950}) after
preliminary reduction (bias correction, flat-fielding, cosmic hits removal,
individual spectrum extraction from the CCD images and its conversion to a
common linear wavelength scale, night-sky line subtraction, correction,
if needed, for the effect of differential atmospheric refraction).

\begin{figure}
\psfig{figure=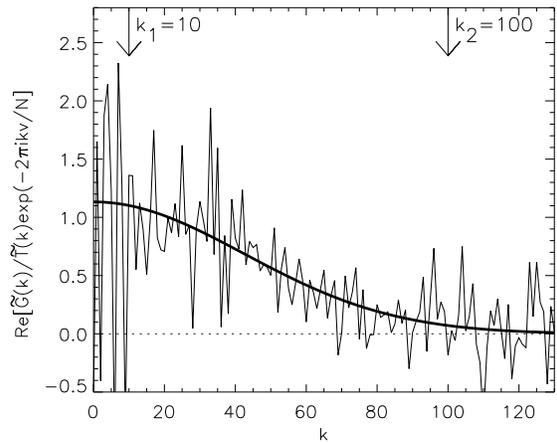,width= 8 cm }
\caption{ Approximation of function (15) by a Gaussian in the Fourier
quotient method. The boundaries of fitting ($k_1$ and $k_2$) are marked.}
\label{ssbsfig}
\end{figure}

After the preliminary reduction the continuum is subtracted from the
spectra. Underestimation of the continuum level leads to appearance of
a low-frequency component in the Fourier spectrum and, as a consequence,
to greater noises for small $k$ in the SSBS method, and of a wide base in the
auto-correlation and cross-correlation peaks in the TD method.
It is not infrequent that high-order polynomials are used to fit a continuum
(Larsen et al., 1983). However, our experience of reduction has shown that
the best result can be achieved if a spectrum smoothed by a median filter
with a window of 150--200 pixels is taken as a continuum. In so doing,
even the broadest absorption spectral features are practically undistorted.

The emission lines in the galaxy spectrum reduce the contrast of the
cross-correlation peak. Depending on the number and the intensity of lines,
different techniques can be employed to remove them. In most cases it
suffices to approximate by a Gaussian the brightest line, determine the
intensity of the rest of the lines with respect to it and subtract from
the spectrum the Gaussians with corresponding weights.

After the continuum subtraction from the emission lines, the spectra are
rebinned onto a logarithmic scale according to (\ref{ton}). To diminish
the effect of ``leakage of energy'' in the Fourier transform due to the
sharp cutoff of the ends of the spectrum (for more details see Brault and
White, 1971; TD, 1979), the spectrum needs to be multiplied by the function
of the form:
$$
F(n)=
\left\{
\begin{array}{ll}
\sin{\frac{\pi n}{2 N p}},&  n<N p \\
1,& N p\le n \le N(1-p) \\
\sin{\frac{\pi (N-n)}{2 N p}},& n>N(1-p),
\end{array}
\right.
$$
here
$p=0.05-0.1$ is the relative size of the spectrum boundaries multiplied
by a sinusoid; examples of such spectra are displayed in Fig.\,\ref{cor2}c,d.

\subsection{Determination of $v$ and $\sigma$}
For each spectrum corresponding to one of the spatial elements
$(x,y)$, relation (\ref{calibe}) is constructed. For this purpose, the
spectrum of the template star $T(x,y)$ is smoothed by Gaussians of different
widths and an estimate of the dispersion is made by the TD method. Because
the cross-correlation peak shape is not always precisely Gaussian, it is
necessary to fix the boundaries of the region of fitting
\begin{figure*}
\psfig{figure=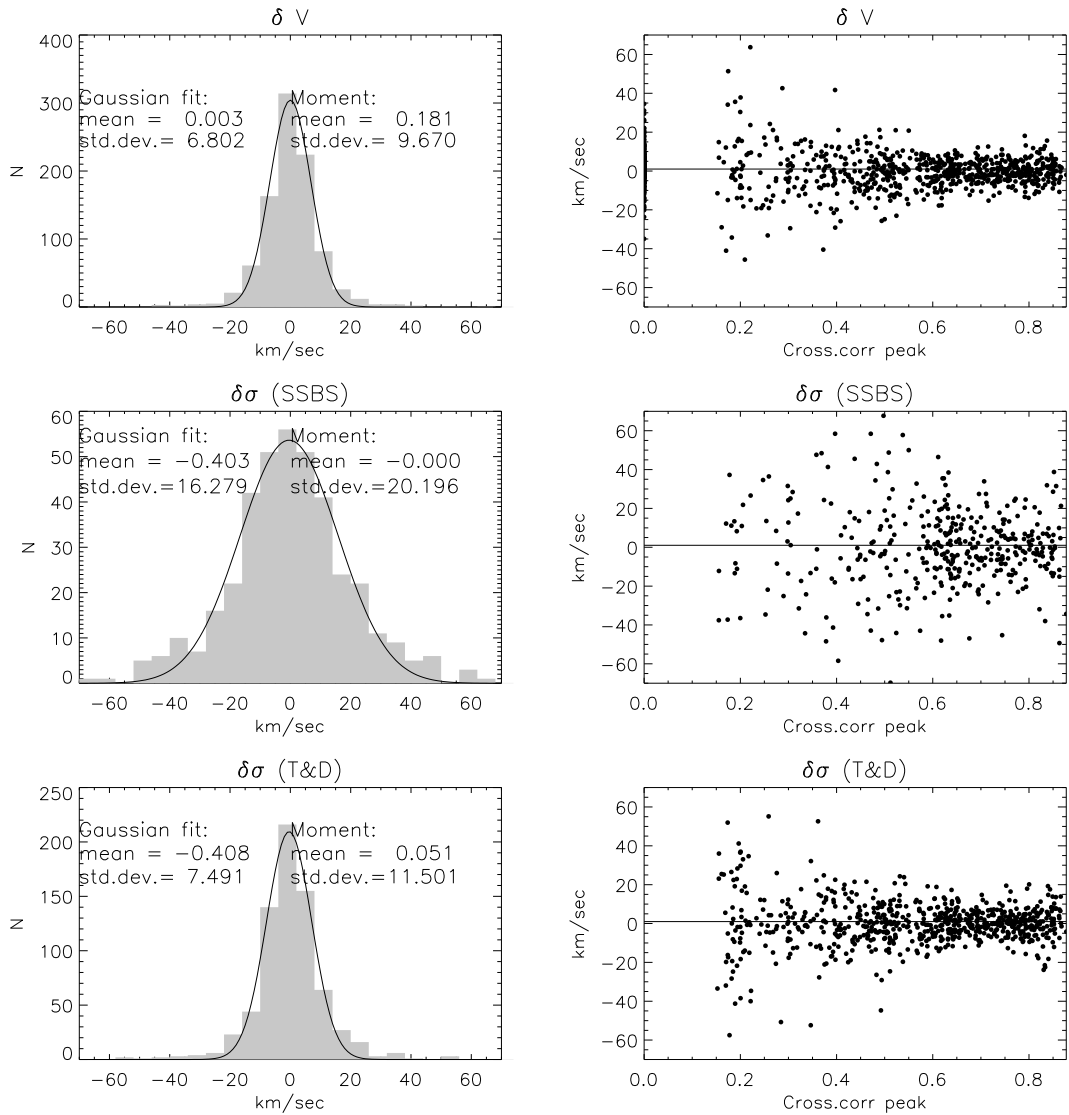,width=15 cm }
\caption{Errors of determination of $v$ (top) and  $\sigma$ (in the middle by
the SSBS method, at the bottom by the TD method) for the galaxy
NGC\,2950 over the whole  MPFS field of view. The errors were estimated from 4
template stars. The left-hand panels are the histogram of distribution of
errors and their Gaussian fit. The right-hand panels show the errors as
a function of cross-correlation peak height.}
\label{dis_err}
\end{figure*}
($[n_o-\Delta n,n_o+\Delta n]$, where $n_o$ is the peak location), in which
a robust fit with a Gaussian is executed, considering that calibration
relation (\ref{calibe}) may change for different  $\Delta n$.

Obtain a set of coefficients:
$\mu(x,y)$, $\alpha_i(x,y),\; i=0,1,2$. If the spread of the coefficients
from point to point is considerable, it is better to approximate these
coefficients along the slit by a second-order polynomial. Construct
cross-correlation functions
$G(x,y)\otimes T(x,y)$ using relation (\ref{corv}), determine $v$, and from
the peak width determine  $\sigma$ with the from of (\ref{td}) and (\ref{calibe2})
(Fig.\,\ref{cor2}). One should use here the same value of $\Delta n$ as the
one used in construction of calibration relation (\ref{calibe}).

To estimate the dispersion by the SSBS method, the right side of expression
(\ref{ssbs2}) is approximated with allowance made for the remarks on the
choice of the limits $k_1$ and $k_2$ (see subsection 2.3 and Fig.\,\ref{ssbsfig}).

By applying the above-described procedure to the spectra of all spatial
elements, produce two-dimensional maps of radial velocities
$v(x,y)$ and velocity dispersions
$\sigma_{TD}(x,y)$, $\sigma_{SSBS}(x,y)$ (the index indicates the method of
construction). The heliocentric velocities (reduced to the centre of the
Sun) are found from the expression:
\begin{equation}
V_{Hel}=v+V_{*}-V_{\odot,*}+V_{\odot,gal}.
\label{vhel}
\end{equation}
Here  $V_{\odot,*}$,   $V_{\odot,gal}$ are the heliocentric velocity
corrections for the template star and the galaxy computed by the standard
method, $V_{*}$ is the radial velocity of a star from the catalogue
(reduced to the centre of the Sun).

\subsection{Choice of spectral range}
The optimum spectral range, in which the analysis described above is made,
must contain absorption lines characteristic of the stellar population of
galactic nuclei, unblended, when possible. It is desirable that the
number of emission lines of ionized gas should be a minimum. Most commonly
the regions around the triplets
MgI\,$\lambda\lambda\,5167, 5172, 5183$\,\AA\, and
CaII\,$\lambda\lambda\,8498, 8542, 8662$\,\AA\, are used. The latter is more
preferable because with one and the same dispersion of the spectrograph
the spectral scale (in km/s) in the region of CaII is 1.6 times as large
as in the region of MgI. For the advantages of using CaII see the paper by
Terlevich et al. (1990).

Unfortunately, the available set of gratings in MPFS made it impossible to work
in the region of the infrared triplet CaII without substantial losses of
light. Experiments performed in the optical range have demonstrated that
the most optimum spectral range
to study kinematics of stars at the centres of SO--Sb galaxies
is $5100-6100$\,\AA.
It includes  strong absorption lines of MgI\,$\lambda\,5175$\,\AA,
FeI\,$\lambda\,5229$\,\AA, FeI+CaI\,$\lambda\,5270$\,\AA, NaI\,$\lambda\,5893$\,\AA\, and others.
At the same time, the emission lines are relatively weak and not numerous,
noticeable mainly only in Sy galaxies ([NI]\,$\lambda\,5199$\,\AA,
[FeVII]\,$\lambda\lambda\,5158, 5276, 5720, 6086$\,\AA, HeI\,$\lambda\,5875$\,\AA).
The observations are made with a grating of  1200\,gr/mm yielding a
reciprocal dispersion of $1.35$\,\AA/px. The spectral range 4800--6100\,\AA\,
contains both the absorption lines described above and the brightest
emission lines  H$_\beta$, [OIII]\,$\lambda\,4959, 5007$, which allows a
simultaneous study of ionized gas kinematics.

\subsection{Assessment of errors}
The assessment of errors in different methods of determination of $\sigma$
and $v$ is considered in some detail in a number of papers (TD, 1979;
SSBS, 1977; Larsen et al., 1983). The influence of random and systematic
errors is, however, best seen when comparing the results obtained from
different template stars during one night. Fig.\,\ref{dis_err} shows the
distribution of errors in the determination of the velocity and velocity
dispersion for all 240 individual spectra of NGC\,2905 in the MPFS field
of view. The mean value for each spatial element was estimated from four
stars of spectral classes G7\,III--K2\,III. The values of the mean errors
estimated by the two methods, as the second moment in the distribution
of deviations of individual measurements from the mean value over the
four stars and as the width of the Gaussian that approximates this
distribution, are indicated beside the histograms of the errors.
These errors are typical of our observations and make
$\delta v=5-10$\,km/s for the velocity,
$\delta\sigma_{TD}=10-15$\,km/s for the velocity dispersion estimated by the
cross-correlation technique, and $\delta\sigma_{SSBS}=15-25$\,km/s for the
Fourier quotient method. For the central regions where the
value of the cross-correlation peak  is $C_{max}=0.8-1.0$, these errors are about
1.5 times as small as the field average. It can be seen from Fig.\,\ref{dis_err}
that the velocity dispersion errors in the SSBS method increase sharply at
$C_{max}<0.7$, that is, one can estimate reliably only the central velocity
dispersion. However, since this method requires no additional calibration
(subsection 2.3), it is good practice to use the SSBS method estimates for
checking the absolute values of the central velocity dispersion.

\begin{figure}
\psfig{figure=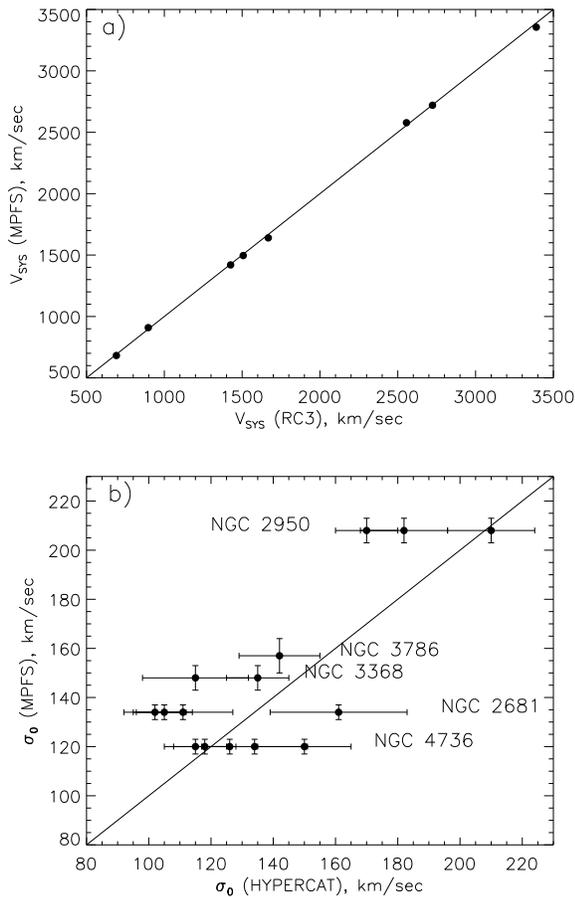,width=8 cm }
\caption{Comparison of systemic velocities (a) and central velocity
dispersions (b) derived from the MPFS observations with the data of the
catalogue  RC3 and the HYPERCAT database.}
\label{compar}
\end{figure}

We have carried out a number of experiments for testing the precision of
the TD and SSBS techniques, using for this purpose the spectra of stars of
class  G6\,III/IV--K3\,III obtained with the MPFS. These spectra were
artificially smoothed and rendered noisy, then they were employed to
estimate the velocity dispersion values. Analysis has shown that if the
signal-to-noise ratio in individual spectra is $S/N > 20$, it does not affect
the accuracy of
$\sigma$ determination, which is  less than $5-10\%$ of the absolute value
of $\sigma$; this is in good agreement with
Fig.\,\ref{dis_err}.

The experiments with smoothing have confirmed the well-known fact that it is
impossible to reliably determine the velocity dispersion if it is smaller
or close to the instrumental profile width. In our case this limit is
70--80\,km/s (see Section 3).

The main factor that affects the accuracy of determination of the dispersion
for a given S/N and spectral interval are the differences in spectral
features of a galaxy and a star which lead to decreasing the contrast of
the cross-correlation peak and to differences in the Fourier spectra of the
SSBS method. When studying the motions of stars in galaxies of early
types, the spectra of red giants G6\,III--K3\,III are normally used as
templates. It is a good idea to take spectra of stars of different spectral
classes from this range during one observing night to select a template
which is the most close to the galaxy spectrum.

For 8 galaxies observed with the MPFS, we compared our measurements of the
systemic velocity (the galaxy centre velocity) with the data of observation
of neutral hydrogen velocities in the 21\,cm line, which had been taken
from the catalogue RC\,3 (Fig.\,\ref{compar}a).
The root-mean-square value of deviations was 17\,km/s without
a systematic shift. Data on the nuclear velocity dispersion of stars were
found in the database
HYPERCAT\footnote{"http://www.obs.univ-lyon1.fr/hypercat"}
and compared with our estimates (Fig.\,\ref{compar}b). No systematic
differences are observed; the great scatter of points is most likely due to
the fact that only using the 2D technique, one can reliably subtract the
spectra from the galaxy dynamical  centre.

\begin{figure*}
\psfig{figure=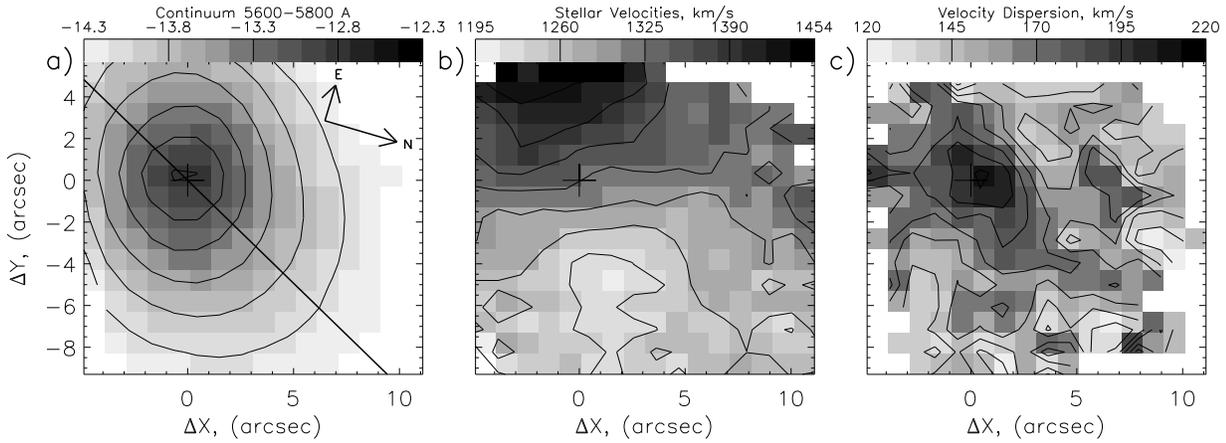,width=16 cm }
     \caption{MPFS maps for the galaxy NGC\,2950. a --- the continuum image
(the surface brightness logarithm is in
$ergs/cm^2/s/\sq''$), the stright line shows the orientation of the main
(outer) bar; b --- the velocity field of stars; c --- the distribution of
the velocity dispersion constructed by the TD method. The origin of
coordinates is at the dynamic centre of the galaxy (shown by cross).}
\label{res2950}
\end{figure*}

\section{Example of data reduction: NGC\,2950}
According to Wozniak et al. (1995) and Friedly et
al. (1996) the SB0 galaxy NGC\,2950 refers to the so-called
double-barred galaxies. In the optical and IR images of the galaxy there is
a bar of about $6''$ in size inside a $38''$ primary bar. The position
angles of the bars are   $PA=332^\circ$ and
$272^\circ$, respectively. NGC\,2950 was observed with the 6\,m telescope
on March 27/28, 2000 using the spectrograph MPFS, the spectral range was
$4820-6200$\,\AA, the dispersion
$1.35$\,\AA/px, the scale $1''$ per lens, the seeing $2''$. The total exposure
was 3600\,s. The field of view of  $16''\times15''$ covered a region of
the inner bar and the central part of the outer bar.

Prior to the continuum subtraction, a smoothing of data was performed in
each spectral bin by a Gaussian of  $FWHM=1''\times1''$ with the aim of
optimum filtering of noises. In so doing the spatial resolution decreased
slightly and was about  2\farcs3. The further processing was conducted
in accordance with Section 4. After the rebinning onto a logarithmic scale,
the velocity bin was 50\,km/s. Five template stars were taken during the
night, 4 of which, yielding a maximum correlation with the galaxy spectrum,
were used to construct average velocity fields and velocity dispersion
(Fig.\,\ref{res2950}). When constructing these fields, the points where
the error was larger than 20\,km/s were masked.

The central velocity dispersion values measured by the TD method
($\sigma=205\pm5$\,km/s) and by the SSBS method ($\sigma=190\pm10$\,km/s)
agree with each other and with the data of HYPERCAT ($170\pm10$\,km/s,
$182\pm14$\,km/s and $210\pm14$\,km/s from different estimates).
The central ellipsoidal structure elongated along $PA\approx330^\circ$,
i.e. the one associated with the ``outer'' bar, is distinguished in the
distribution of velocity dispersion (Fig.\,\ref{res2950}c). At the present
time, we continue the work over the study of kinematics of double-barred
galaxies; preliminary results are available in the paper by Moiseev and
Afanasiev (2001).

\section{Conclusions}
With the above-described procedure one can construct velocity fields and velocity
dispersions of stars in the central parts of early-type galaxies on the
basis of data obtained with the spectrograph MPFS. The accuracy of velocity
measurement is on the average
$\delta v=5-10$\,km/s, that of velocity dispersion measurement is
$\delta\sigma_{TD}=10-15$\,km/s and $\delta\sigma_{SSBS}=15-25$\,km/s
for TD and SSBS methods, respectively. This technique allows us to take account of the
variations of the instrumental profile and produce reliably
two-dimensional maps of the velocity dispersion if it is not less than
70--80\,km/s. The systemic velocities and the central velocity dispersions
that we have measured in 8 S0--Sb galaxies and the HYPERCAT data do not
show systematic differences.

We associate the future prospects of studying kinematics of stars in galaxies
aided by the spectrograph MPFS with the possibility of observing in the region
of the infrared triplet CaII. The accuracy of measurement of velocities and
velocity dispersion must in this case be about 1.5--2 times better.

\begin{acknowledgements}
The author thanks V.L.\,Afanasiev for formulation of the problem and
discussion of results. In the course of work the database HYPERCAT being
developed at the Observatoire de Lyon was used.
\end{acknowledgements}

\newpage
\end{document}